# Manganese-based macrocyclic chelates as novel MRI contrast agents: *In vivo* imaging in a porcine model

Pål B. Marthinsen, MD*, Tuva R. Hope, PhD*, Wibeke Nordhøy, PhD*, Deirdre B. Cassidy, PhD †, Adrian P.L. Smith, PhD †, Paul M. Evans, PhD†, Atle Bjørnerud, PhD *‡

* Division of Radiology and Nuclear Medicine, Oslo University Hospital, Oslo, Norway
† GE Healthcare Pharmaceutical Diagnostics, Chalfont St Giles, UK
‡ Department of Physics, University of Oslo, Oslo, Norway

Conflicts of interest: D.B.C, A.P.L.S and P.M.E are employees of GE Healthcare.
The study was supported by a grant from the Norwegian Research Council (269810/O20)

Corresponding author e-mail: atlebjo@fys.uio.no

## Abstract

**Objectives:** Manganese (Mn) based MRI contrast agents (MBCAs) have recently been proposed as alternatives to the currently used class of low molecular weight chelates of the paramagnetic metal gadolinium (Gd). Unlike the exogenous metal Gd, Mn is an endogenous paramagnetic metal with known biochemical pathways in the human body for excretion and metal regulation, which may alleviate the raised concerns about the safety of existing GBCAs. The aim of this study was to investigate the distribution, kinetics and image enhancement properties of a class of novel Mn-based macrocyclic chelates in a naïve porcine model

**Methods:** Three novel macrocyclic MBCAs, AH114608, GEH300017 and GEH200486, were tested and compared to gadoterate meglumine. Twelve female adult pigs were divided into four groups (n=3 for each CA). At 3 T MRI, T1 relaxometry analysis and T1-weighted enhancement properties were measured longitudinally in liver, kidney, blood pool in the left ventricle and descending aorta and the myocardium at five timepoints 30 minutes apart following CA administration and one timepoint prior to CA administration. MR angiography was performed during the CA bolus phase and arterial maximum intensity projections covering descending aorta and kidneys were generated. CA kinetics was estimated from analysis of plasma CA concentrations by inductively coupled plasma optical emission spectrophotometry (ICP-OES).

**Results:** All four CAs exhibited T1-enhancing properties in the blood pool with GEH200486 having the largest increase in T1 relaxation rate (R1), GEH300017 and gadoterate meglumine having similar R1 increase and AH114608 a lower peak R1 change. A persistent increase in liver, kidney and myocardium R1 (up to about 210 minutes post CA injection) was observed with AH114608. To a lesser extent, a persistent increase of liver enhancement was also observed in T1-weighted images for GEH300017 and GEH200486 compared to gadoterate meglumine. All four CAs had similar bi-exponential plasma kinetics characterized by a rapid distribution phase and a slower elimination phase.

**Discussion/Conclusion:**

Based on relaxometry and visual assessment of T1 enhancement properties at 3 T, we have identified MBCA candidates, AH114608, GEH300017 and GEH200486 with predominantly renal clearance. GEH300017 and GEH200486 showed comparable efficacy in terms of vascular T1 relaxation, and comparable to the reference GBCA. AH114608 demonstrated a different *in vivo* profile with a lower peak enhancement in the blood and prolonged liver, kidney and myocardium enhancement. The T1-enhancing properties of these novel Mn macrocyclic CAs can be used with routine clinical protocols and could be potentially utilised as an alternative to GBCAs for contrast-enhanced MRI procedures.

# Introduction

The use of contrast agents (CA) has proven its merit in improving the diagnostic quality of magnetic resonance imaging (MRI) and every year CAs are used in approximately 50 million MRI procedures to enhance the detection and monitoring of pathological processes (1). All commercially available MRI CAs today are low molecular weight chelates of the paramagnetic metal gadolinium (Gd). In 2006, a direct association between nephrogenic systemic fibrosis (NSF), a serious condition characterized by fibrosis of skin, joints, eyes, and internal organs, and the administration of Gd-based CAs (GBCAs) in patients with severely impaired kidney function, was found (2). Recently, reports have emerged regarding Gd retention in the brain following repeated injections of GBCAs to patients with both normal and reduced kidney function (3-7). The clinical significance of the accumulation of Gd in brain tissue (dentate nucleus and globus pallidus), is currently unknown. However, uncertainty regarding the safety of using GBCAs has been ignited based on perceptual and scientific factors, since both NSF and brain retention have been hypothesized to be linked to the stability of GBCAs (2, 8).

The human body has no specific biochemical pathways for handling exogenous Gd and replacing Gd with an MRI active metal naturally occurring in the body may alleviate the raised concerns. Manganese (Mn) is paramagnetic endogenous metal, mainly found in bone, brain, kidney, and liver, is an essential cofactor for several intracellular activities, and the human body has biochemical pathways for excretion and regulation of Mn (9). Two Mn agents, mangafodipir trisodium (Teslascan, Nycomed, Oslo, Norway), a safe and efficacious hepatobiliary contrast agent (10, 11), designed to release $Mn^{2+}$ after injection, and an oral gastrointestinal contrast agent containing Mn(II) chloride (LumenHance, Bracco Diagnostics, Milan, Italy) were clinically approved but were withdrawn from the market for commercial reasons (12-14). Due to more challenging coordination chemistry compared to GBCAs, the development of a sufficiently stable general-purpose Mn-chelate with suitable MR contrast enhancement properties has been more complex, and data regarding Mn complexes are still incomplete compared to Gd complexes (15). There are no Mn-based CAs (MBCAs) available for clinical use today.

Two recently published studies report results from a newly developed linear MBCA, Mn-PyC3A which was tested for efficacy in two animal models. Comparable contrast enhancement to GBCA on both MR angiographic and T1 weighted acquisitions was found (16, 17), suggesting that MBCAs are feasible alternatives to GBCAs in terms of image enhancement efficacy.

The purpose of this study was to evaluate the distribution, kinetics, and image enhancement properties of three novel macrocyclic Mn-chelates in a naïve porcine model and compare these to a clinically approved GBCA using suitable MR imaging protocols and image analysis methods. The study protocol was specifically designed to map T1 relaxation times in various body regions across time.

# Materials and Methods

## Ethical Considerations

The study protocol was approved by the Norwegian Food Safety Authority and conducted in accordance with the Norwegian Regulation on Animal Experiment (FOR-2015-06-18-761). Animal handling was performed by personnel with certified training required for working with research animals.

## Contrast Agents

Three novel MBCAs, AH114608, GEH300017, and GEH200486, developed and supplied by GE Healthcare (Chalfont St Giles, UK), were formulated in water for intravenous injection to a concentration of 0.5 mmol/mL to be administered at 0.2 mL/kg to give a final dose of 0.1 mmol/kg. The molecular structures of the three novel Mn complexes are given in the patent WO2017220610. All MBCAs are macrocycle based chelates where A114608 is the core compound structure with no extra substituents. GEH300017 and GEH200486 are derivatives of the basic compound comprising side arms for structural diversity. A GBCA control, gadoterate meglumine (Dotarem, Guerbet, Villepinte, France) was administered at the same dose and volume as the MBCAs. The $r_1$ and $r_2$ relaxivities (units $mM^{-1}s^{-1}$) in male human serum at 37°C and 300 MHz (3 T) of the three MBCAs were determined by NMR relaxometry (Stelar SMARTracer NMR console equipped with a VTC90 variable temperature control unit and running AcqNMR v95 software) to be r1 [2.1, 3.7, 5.0] and r2 [8.2, 16.2, 18.9] for AH114608, GEH300017, and GEH200486, respectively (18).

## Animals and Interventional Procedures

A total of 13 naïve female pigs (48kg ± 5kg, age 4-5 months) were included in the study. One pig was used for piloting and protocol optimization and the remaining 12 were subdivided into four groups (n=3 for each CA). Details can be found in Table 1. Upon arrival, premedication mixture of 20 mg/kg ketamine hydrochloride (Ketalar, Pfizer, Brussels, Belgium), 3 mg/kg azaperone (Stresnil, Janssen, Beerse, Belgium) and 0.02 mg/kg atropine sulphate (Atropin, Takeda, Tokyo, Japan), was injected intramuscularly in the neck region through a 100 cm soft-walled tubing. This allowed the animal to move freely during the administration of pre-medication drugs.

*Table 1 Study animals, mean weight and contrast agents*

| Pig ID | Weight (kg) | CA administered |
|---|---|---|
| 0, 7, 8 | 48.3 kg ± 1.9 | Gadoterate meglumine |
| 1, 2, 6 | 47.7 kg ± 6.3 | GEH300017 |
| 3, 4, 5 | 50.0 kg ± 8.0 | AH114608 |
| 9,10,11 | 44.8 kg ± 2.0 | GEH200486 |

Venous access was gained by cannulation of the auricular vein, and additional 1-2 mg pentobarbital sodium (Exagon, Richter Pharma AG, Wels, Austria) could be given intravenously if required during transportation to the imaging facility. General anaesthesia was used for the remaining part of the procedure. Intravenous (IV) infusion of 5-6 ml/kg/h ringer acetate, 1-2 mg/kg morphine hydrochloride (Morfin, Takeda, Tokyo, Japan), and 1-3 mg/kg pentobarbital sodium was given prior to tracheostomy. The anaesthesia was maintained by mechanical ventilation with 1-2% isoflurane and a parallel IV infusion of 1-2 mg/kg/h morphine. All pigs were fitted with an external jugular vein catheter for administration of CA. An additional catheter was inserted in the carotid artery, for hemodynamic monitoring throughout the procedure and for drawing of blood samples to estimate creatinine levels and Mn/Gd plasma concentration. Peripheral oxygen saturation, $CO_2$ levels, body temperature, anaesthesia depth and diuresis were continuously checked. Anaesthetic personnel were present during the full experiment. Immediately after the completion of the final imaging session (approximately at 225 minutes after the CA administration), the animals were euthanized with 1g IV thiopental sodium (Pentocur, Abcur AB, Helsingborg, Sweden) followed by 100 mmol IV potassium chloride (KCl).

## Blood Sample Analysis

Arterial blood samples were drawn from the carotid artery and collected into 3.5 mL vacuum tubes at baseline and at four time points, at approximately 2, 30, 90, and 180 minutes after CA injection. After withdrawal, heparine sodium (Heparin, LEO Pharma, Ballerup, Denmark) was mixed with each blood sample before the mixture was frozen at -20 °C and analyzed within 30 days. The concentration of Mn and Gd was measured by inductively coupled plasma optical emission spectrophotometry (ICP-OES). Using the exact timestamp of blood sampling for each animal, plasma Mn and Gd concentration versus time were fitted to the biexponential function (19):

$$[CA] = A \left( f1 \cdot exp\left(\frac{-t}{T_{1/2\ fast}}\right) + f2 \cdot exp\left(\frac{-t}{T_{1/2\ slow}}\right) \right) \quad (1)$$

where [CA] is the molar concentration of Mn or Gd, $t$ is the time since CA injection, A is a scaling constant, f1 and f2 are the relative fractions of the half-lives of the fast ($T_{1/2\ fast}$) and slow ($T_{1/2\ slow}$) time components, respectively and $f1+f2=1$. Curve fitting was performed in Matlab R2019 (The MathWorks, Inc, USA).

Whole blood creatinine levels, as a marker of kidney function, were measured at all time points.

## Magnetic Resonance Imaging

MRI scanning was performed on a Philips Ingenia 3.0 Tesla scanner (Philips Healthcare, Best, the Netherlands) approved for animal use. All pigs were imaged with a body coil under full anaesthesia. Table 2 summarizes the scan parameters for the different MRI sequences used. The main protocol was performed prior to administration of CA and then repeated every 30 minutes, with the last session starting 210 minutes after CA injection. Breath-hold was applied for the THRIVE 3D GRE and the 3D MRA sequences. 3D MRA was acquired during the first-pass phase using the Philips 'BolusTrak' technique (20). With this, rapid low-resolution images are acquired during bolus injection allowing visual observation of bolus arrival in the upper part of the descending aorta to aid in correct timing of the MRA acquisition. A CA dose of 0.1 mmol Mn or Gd /kg b.w. was administered IV using a power injector (Medrad Spectris Solaris) at a flow rate of 3 mL/s. This was followed by flushing with 20 mL saline solution to clear any remaining contrast agent from the connecting tube.

*Table 2. Imaging protocol and sequence parameters*

| Sequence type | | Parameters | |
|---|---|---|---|
| 3D Look-Locker (T1 mapping) | Orientation | Sagittal | Repeated every 30 minutes |
| | TR | 2.3 ms | |
| | TE | 1.1 ms | |
| | TI/TD | 47 ms / 3957 ms | |
| | Flip angle | 4° | |
| | Voxel size | 1.88x1.88x4 mm$^3$ | |
| | Duration | 336 s | |
| 3D Look-Locker (T1 mapping) | Orientation | Axial | |
| | TR | 2.2 ms | |
| | TE | 1.2 ms | |
| | TI/TD | 47 ms/ 3957 ms | |
| | Flip angle | 4° | |
| | Voxel size | 1.79x1.79x4 mm$^3$ | |
| | Duration | 336 s | |
| THRIVE: 3D GRE | Orientation | Axial | |
| | TR | 3.6 ms | |
| | TE | 1.76 ms | |
| | Flip angle | 10° | |
| | voxel size | 1x1x4 mm$^3$ | |
| | duration | 12.4s | |
| MRA | Orientation | Coronal | Once, during bolus injection |
| | TR | 5.4 ms | |
| | TE | 1.3 ms | |
| | Flip angle | 35° | |
| | Voxel size | 0.7 x 0.7 x 2 mm$^3$ | |
| | Duration | 26 s | |

### Image Analysis

T1 maps were generated from the 3D Look-Locker sequence. Here, the modulus signal following a 180-degree inversion pulse was fitted to the inversion recovery signal equation using a non-linear least-squares algorithm, yielding pixel-wise estimates of T1 relaxation times and relaxation rates (R1=1/T1) and the corresponding residual root mean square error (RMSE) of the curve fit (21). The Look-Locker sequence produced a total of 56 separate inversion times (TIs) equally spaced between 47 ms and 3957 ms and the sequence was repeated three times to give three independent data points at each TI-value.

R1 values were extracted from a single elliptical region of interest (ROIs) manually placed in the left ventricular myocardium, the left ventricle blood pool, the right kidney parenchyma and the descending aorta. In the liver, R1 from three separately measured ROIs were averaged. The ROIs were carefully chosen in regions with homogenous signal and low RMSE. Renal parenchyma ROIs were obtained from axial R1 maps and sagittal R1 maps were used for the remaining ROI analysis. Figure 1 shows examples of ROI placements in the sagittal and axial R1 maps.

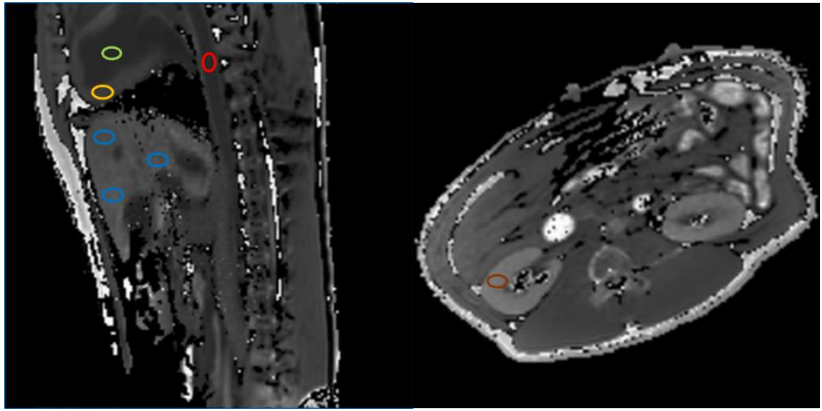

**Figure 1** *Sample case showing sagittal (left) and axial (right) R1 maps and corresponding ROI placements for descending aorta (red, left image), left ventricle blood pool (green), myocardium (yellow), liver (blue, average of three ROIs) and kidney (red, right image).*

To better estimate possible myocardial CA retention effects, additional analysis was performed where the ratio of myocardial R1 change was normalized to the same R1 change in the left ventricle blood pool. This ratio is expected to remain constant over time if the tissue distribution of the CA is confined to the vascular and extravascular space with equal excretion rate in a steady-state condition. If, on the other hand, the agent is retained in tissue, the ratio is expected to increase over time.

To compare visual liver enhancement across animals and CAs, the THRIVE T1-weighted images were normalized to the intensity measured in a ROI in the paraspinal muscle. Finally the ratio of the normalized liver signal intensity pre/post-contrast (last time point) was calculated for all four agents.

Visual evaluation of vascular enhancement was assessed from maximum intensity projections (MIPs), generated from the MRA data acquired during the CA bolus phase. Window level/width of the MIPs was adjusted to ensure comparable vascular appearance for all CAs. Additional quantitative analysis of contrast enhancement was performed by measuring, in the raw MRA data, the signal intensity in the descending aorta normalized to the intensity in the paraspinal muscle. The average value from three ROIs placed at different positions along the aorta was used for the quantitative enhancement estimation.

All image processing was performed in nordicICE (NordicImagingLab, Bergen, Norway), except MIP reconstructions which were generated in ImageJ (Rasband WS. ImageJ, U.S. National Institutes of Health, Bethesda, Maryland, USA, imagej.nih.gov/ij/, 1997—2012.)

### Statistical Analysis

Descriptive statistics are reported as ROI mean and standard deviation across animals in each CA group. Group differences are reported as visual or quantitative observations since the study was not powered for statistical inference of group differences.

## Results

The imaging protocol was successfully completed in all 12 animals, without any adverse reactions observed. The vital signs were stable for each pig across the course of the experiment, before and following injection of CA. One animal in the GEH300017 group (ID 2) had to be excluded from the MRA analysis due to delayed start of MRA acquisition relative to the arterial bolus arrival time, causing poor arterial enhancement.

### Whole Blood Analysis

The creatinine blood levels showed a trend towards a slight increase in creatinine levels towards the end of the procedures. This was independent of the CA administered (including GBCA or MBCAs) and is shown in Figure 2.

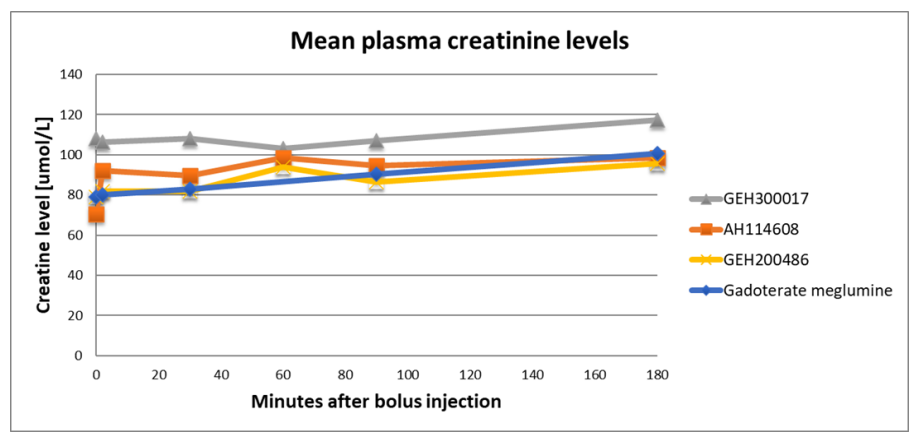

**Figure 2.** *Time-dependent mean plasma creatinine levels for pigs receiving different CAs. N=2 for gadoterate meglumine and n=3 for the other CAs.*

The results of the bi-exponential curve fit to Equation 1 are shown in Table 3. Figure 3 shows the plasma concentration-time curves for the different agents, including the corresponding curve fits. All agents exhibited a bi-exponential kinetic and there was a trend towards longer half-lives for both the fast and slow components for AH114608 and GEH300017 compared to the two other agents.

**Table 3.** *Bi-exponential plasma kinetic parameters from ICP analysis, where $T_{1/2}$ fast (alpha) = distribution phase (movement from blood to ECF) and $T_{1/2}$ slow (beta) = elimination phase (removal from blood via renal excretion).*

| Contrast agent | Fast fraction | $T_{1/2}$ fast (minutes) | $T_{1/2}$ slow (minutes) |
|---|---|---|---|
| AH114608 | 0.77 | 12.0 | 141.6 |
| GEH300017 | 0.65 | 14.4 | 136.7 |
| GEH200486 | 0.71 | 7.3 | 101.8 |
| Gadoterate meglumine | 0.62 | 9.2 | 107.0 |

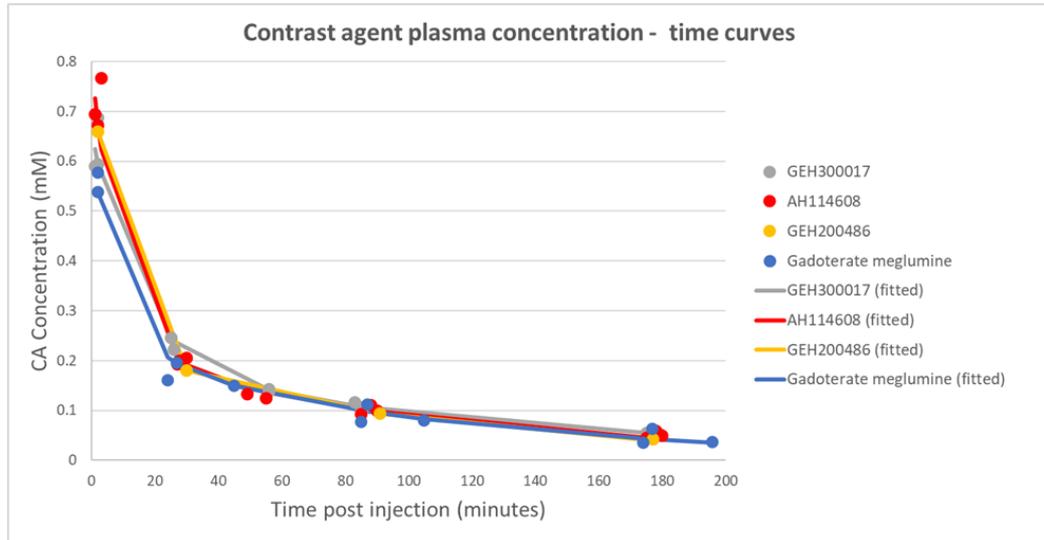

**Figure 3.** *Plasma concentration of Mn/Gd from ICP for the four test agents, including bi-exponential curve fits. The corresponding kinetic parameters are given in Table 3.*

## Relaxometry Results

Figure 4 shows measured R1 changes over time for the target tissues liver parenchyma, descending aorta, left ventricle, and kidney cortex. Similar results for myocardium are shown in Figure 5 (a). The average results are plotted as ΔR1 = R1(t)-R1(0) for each CA (n=3) and ROI separately. GEH200486 exhibited the largest peak change in R1 in all target tissues whereas GEH300017 and gadoterate meglumine gave similar peak ΔR1 changes. AH114608 exhibited a lower peak R1 change in all tissues. Further, AH114608 showed a marked increase in liver R1 over time post-injection, and to a lesser extent also in kidney and myocardium. A similar time-dependent increase in R1 was not observed with the other two MBCAs or with gadoterate meglumine. Figure 5 b) shows time curves of the change in myocardial R1 normalized to the corresponding change in left ventricle blood pool. The linear increase in this ratio for AH114608 suggests prolonged myocardial retention of this agent.

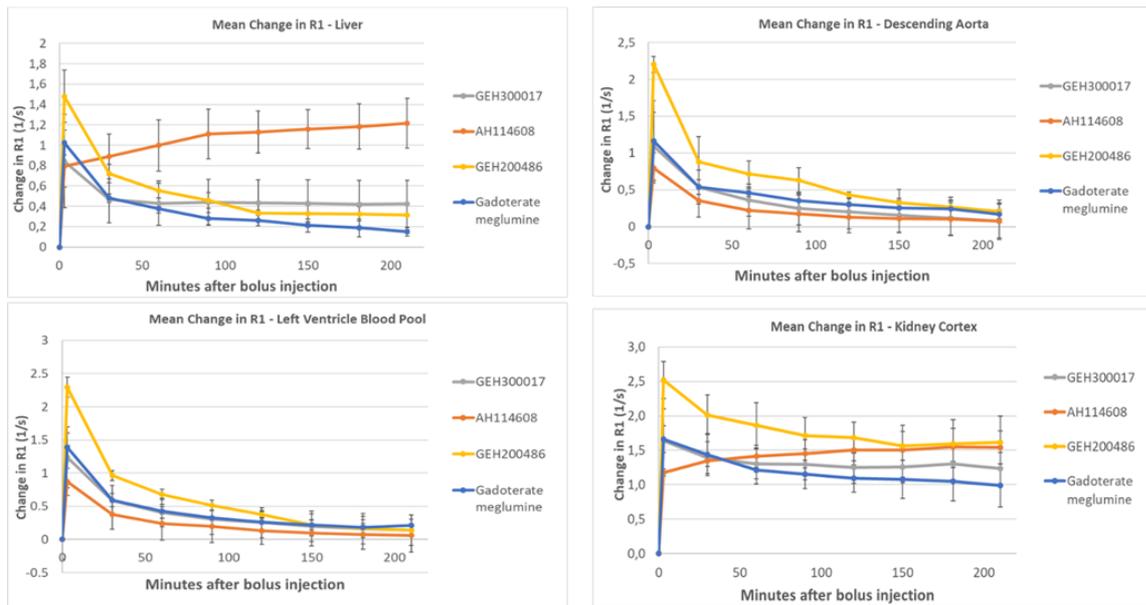

**Figure 4.** *Time-dependent change in R1 relaxation rates from pre-contrast R1 for the different target organs and contrast agents.*

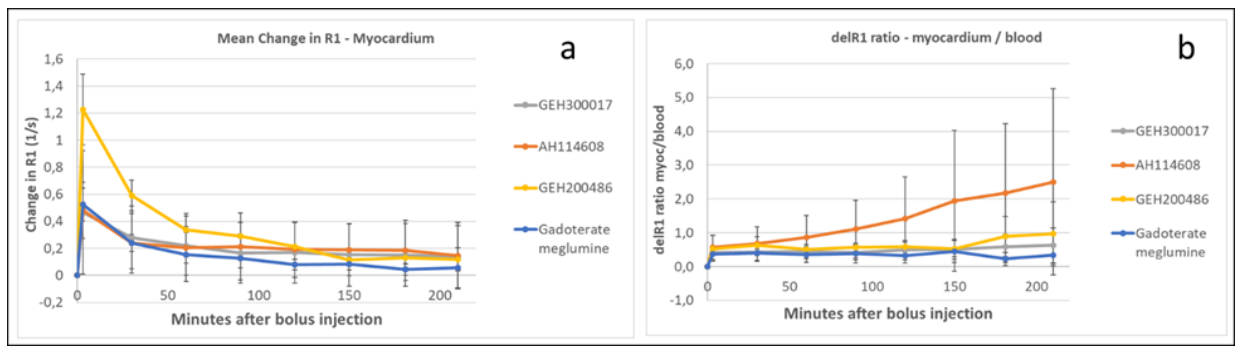

**Figure 5.** *(a) Time-dependent change in myocardial relaxation rates R1 compared to pre-contrast R1. (b) Time-dependent evolution of the ratio of R1 relaxation rates in myocardium/blood indicating myocardial retention for AH114608.*

## Visual T1-w Liver Enhancement

Figure 6 shows sample cases of the intensity normalized 3D T1-weighted THRIVE series of the liver pre - and approx. 210 minutes post-CA administration. There is visible late post contrast enhancement for all three MBCAs, and most pronounced for AH114608; in agreement with measured R1 relaxation rate changes. The corresponding measured ratios of normalized signal intensity pre/post-contrast are shown in Figure 7, confirming the elevated post-contrast liver intensity, especially for AH114608.

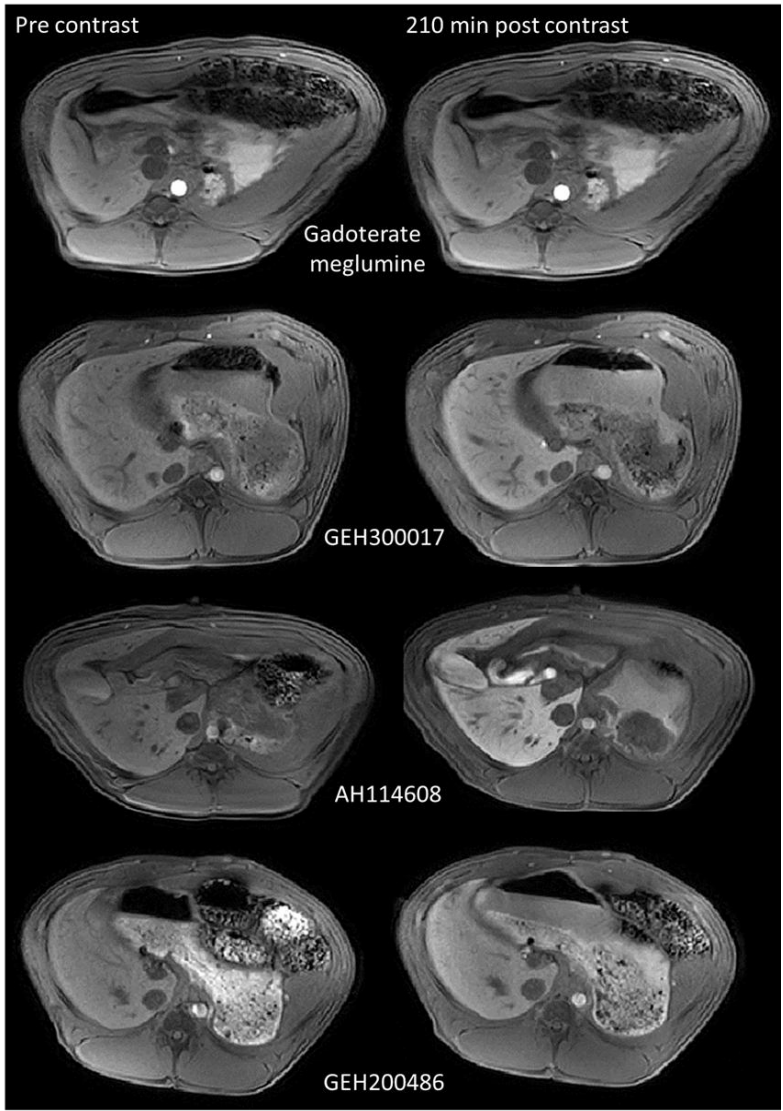

**Figure 6.** *Sample cases showing intensity normalized (to paraspinal muscle) liver images pre-contrast administration (left) and approximately 210 minutes post-contrast administration (right) for all four agents.*

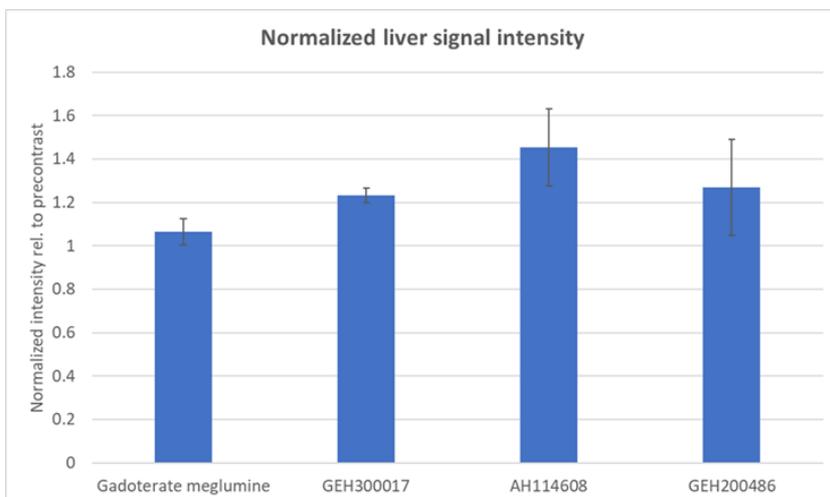

**Figure 7.** *Ratio of normalized signal intensity in liver at 3 hours post-contrast relative to pre-contrast normalized intensity.*

## MR Angiography

Figure 8 shows sample MIPs obtained from one animal in each CA group. All three MBCA and GBCA agents provided good vascular contrast enhancement and no clear visible difference in the enhancement or quality of the MIPs between any of the agents could be observed beyond the technical limitations of the timing of the bolus. The result of the analysis of relative enhancement in the descending aorta relative to paraspinal muscle is shown in Figure 9, indicating a lower relative enhancement in scans with AH114608 compared to the other three CAs, in agreement with the lower peak change in R1 for this agent.

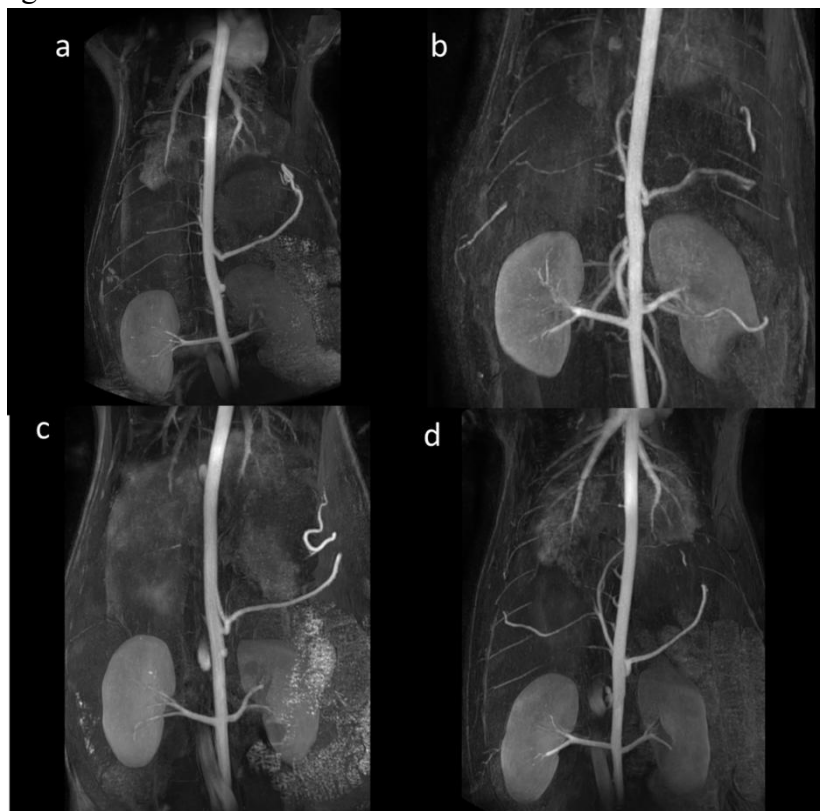

**Figure 8.** *Sample maximum intensity projection angiograms acquired during the arterial bolus phase for gadoterate meglumine (a), GEH300017 (b), AH146018 (c), and GEH200486 (d).*

## Discussion

The main purpose of this study was to investigate the kinetics and imaging efficacy of a new class of extracellular macrocyclic MBCAs compared to a reference GBCA in an adult porcine model.

The three MBCAs investigated differ in chemical structure and relaxivity properties, but all three agents have shown to exhibit relaxivity properties suitable for T1-enhancing agents at clinical field strengths. The MBCAs have comparable $r_1$ to GBCAs but a significantly higher transverse relaxivity, $r_2$, due to the chemistry of the chelate and the parametric properties of Mn (18). Although the number of animals in each test group was too small to make statistical inference, GEH300017 exhibited the highest peak change in R1 relaxation rate and AH114608 the lowest peak R1 whereas GEH200486 had enhancement properties similar to the reference GBCA. The relative scale of signal enhancement is in agreement with the measured *in vitro* T1 relaxivities of these agents (18).

A late and prolonged elevated R1 relaxation rate in the liver was observed for the AH114608 group (Figure 4). As seen in Figure 6, this enhancement effect was also clearly visible in the T1-weighted THRIVE series, comparing pre-contrast to last imaging time point, in the pig administered with AH114608 but not in the pig administered with gadoterate meglumine. Similar enhancement patterns were seen in all animals in each group (data not shown). Minimal late enhancement, less than that observed with AH114608, was also noticed for GEH300017 and GEH200486, both in terms of visual enhancement (Figure 6) and elevation in liver R1 (Figure 4). The evidence of prolonged and persistent liver T1-enhancement observed with AH114608 may suggest a contribution from the transport of free Mn through the hepatobiliary system, a known clearance route. Similar findings, but to a lesser extent, were found in the myocardium for this agent. These findings warrant further studies including investigation of the effect of addition of excess ligand to the formulation.

Despite the differences in quantitative R1 changes, all Mn-based test substances exhibited good T1-efficacy in the MR angiography series acquired during the bolus phase. There was no clear visible difference in the quality of enhancement in the resulting MIPs (Figure 8). It should, however, be noted that the magnitude of the vascular enhancement in an MRA experiment is critically dependent on the timing of the acquisition relative to the passage of the bolus, making a direct comparison between separate acquisitions difficult. The quantitative assessment measured from the MRA data of vascular enhancement in the descending aorta (normalized to paraspinal muscle), did indicate a higher degree of enhancement for GEH200486 and less enhancement for AH114608 compared to GEH300017 and gadoterate meglumine, in agreement with the relaxometry results.

The study has limitations, the low number of animals (n=3) in each contrast agent group provided insufficient power for statistical inference of differences in contrast agent efficacy. This is a common limitation in studies of this nature using large animal models. The results obtained in the current *in vivo* study are, however, in agreement with recent *in vitro* relaxometry results obtained with the same MBCAs investigated here. In this phantom study, GEH200486 was found to exhibit a higher $r_1$ relaxivity and AH114608a lower $r_1$ relaxivity compared to a gadoterate meglumine, measured on a 3 Tesla clinical scanner. GEH300017 had similar $r_1$ relaxivity to the GBCA in this study (18).

All Mn agents and the comparative GBCA were found to exhibit bi-exponential kinetics (Figure 3 and Table 3), as expected for small molecular weight, renally excreted extracellular fluid (ECF) agents (19). Slightly longer time constants were observed for AH114608 and GEH300017 compared to the GE200486 and gadoterate meglumine. However, given that only four time-points were available for the analysis, these results should be confirmed in more detailed plasma kinetics studies including more data points, especially at the early time-points after CA administration.

Prolonged T1-enhancement was detected in liver for all MBCAs and in the myocardium for AH114608. Later timepoints (beyond 210 minutes) were not investigated to assess the liver wash-out kinetics of Mn. Possible enhancement in other organs (e.g. the pancreas) was not investigated due to sub-optimal visualization in the acquired image series.

In conclusion, based on relaxometry and visual assessment of T1 enhancement properties, we have identified MBCA candidates, AH114608, GEH300017 and GEH200486 with predominantly renal clearance as potential alternatives to existing general-purpose GBCAs. GEH300017 and GEH200486 showed efficacy in terms of vascular T1 relaxation, comparable to the reference GBCA. AH114608 demonstrated a different *in vivo* profile with a lower peak enhancement in the blood and prolonged liver enhancement. Despite having higher $r_2$ relaxivities, these novel macrocyclic MBCAs acted as T1 contrast agents using routine clinical protocols and can potentially offer an alternative to a GBCA for contrast-enhanced MRI procedures.